\documentclass{INTERSPEECH2023}

\interspeechcameraready

\usepackage{amsmath,graphicx}
\usepackage{booktabs}
\usepackage{multirow}
\usepackage{amsmath}
\usepackage{amsfonts}
\usepackage{mathtools}
\usepackage{xurl}

\newcommand{\etal}{\textit{et al}~}

\title{Semi-supervised learning for continuous emotional intensity \\ controllable speech synthesis with disentangled representations}
\name{Yoori Oh$^{1}$, Juheon Lee$^{2}$, Yoseob Han$^{3}$, Kyogu Lee$^{1,4,5}$}
\address{$^1$Department of Intelligence and Information, Seoul National University \\
     $^2$Supertone, Inc.  \quad 
	$^3$School of Electronic Engineering, Soongsil University \\
    $^4$Interdisciplinary Program in Artificial Intelligence, Seoul National University \\
	$^5$Artificial Intelligence Institute, Seoul National University 
	} 
\email{\footnotesize\{yoori0203, kglee\}@snu.ac.kr, juheon@supertone.ai, yoseob.han@ssu.ac.kr  }

\begin{document}

\maketitle

\begin{abstract}
Recent text-to-speech models have reached the level of generating natural speech similar to what humans say. But there still have limitations in terms of expressiveness. The existing emotional speech synthesis models have shown controllability using interpolated features with scaling parameters in emotional latent space. However, the emotional latent space generated from the existing models is difficult to control the continuous emotional intensity because of the entanglement of features like emotions, speakers, etc. In this paper, we propose a novel method to control the continuous intensity of emotions using semi-supervised learning. The model learns emotions of intermediate intensity using pseudo-labels generated from phoneme-level sequences of speech information. An embedding space built from the proposed model satisfies the uniform grid geometry with an emotional basis. The experimental results showed that the proposed method was superior in controllability and naturalness. 
\end{abstract}

\section{Introduction}
Synthesized speech from deep learning-based text-to-speech (TTS) models \cite{1_tan2022naturalspeech,2_kim2021conditional,3_ren2020fastspeech} have already shown excellent performance about naturalness. It is suitable and sufficient for general information delivery purposes to apply a speech synthesis system to real-world applications. However, it is difficult to synthesize expressive speech including paralinguistic characteristics such as pitch, stress, tone, and rhythm.

Expressive speech models are increasingly necessary, so emotional TTS research is being aggressively pursued. There are several works \cite{4_wang2018style,5_wu2019end,6_cai2021emotion,7_lee2017emotional,8_tits2019exploring,9_li2021controllable,10_li2021controllable,11_um2020emotional,12_im2022emoq,zhu2019controlling,lei2021fine,lei2022msemotts} related to emotional speech synthesis model. First, some studies \cite{4_wang2018style,5_wu2019end,6_cai2021emotion} proposed methods to extract emotional information from reference speech. Global style token (GST) \cite{4_wang2018style} demonstrated a style encoder trained by unsupervised learning to extract style embedding vector from reference speech and then exploited it to synthesize emotional speech. Other studies \cite{5_wu2019end,6_cai2021emotion} used a speech emotion recognition (SER) model to learn a speech emotion embedding space. Authors \cite{7_lee2017emotional,8_tits2019exploring} proposed a method to utilize categorical emotion labels. Specifically, Lee \etal \cite{7_lee2017emotional} applied the emotion labels to the attention RNN to enable emotional speech synthesis. Tits \etal \cite{8_tits2019exploring} fine-tuned a pretrained speech synthesis model with a small set of emotional dataset. Unfortunately, speech synthesized by the previous methods \cite{4_wang2018style,5_wu2019end,6_cai2021emotion,7_lee2017emotional} provided only a coarse-grained expression because the entire sentence has been adjusted with one global information. Therefore, it is difficult to reflect the user's requirements for fine-grained control in the emotional TTS model.
\begin{figure}[t!]
	\centering
	\includegraphics[width=0.65\linewidth]{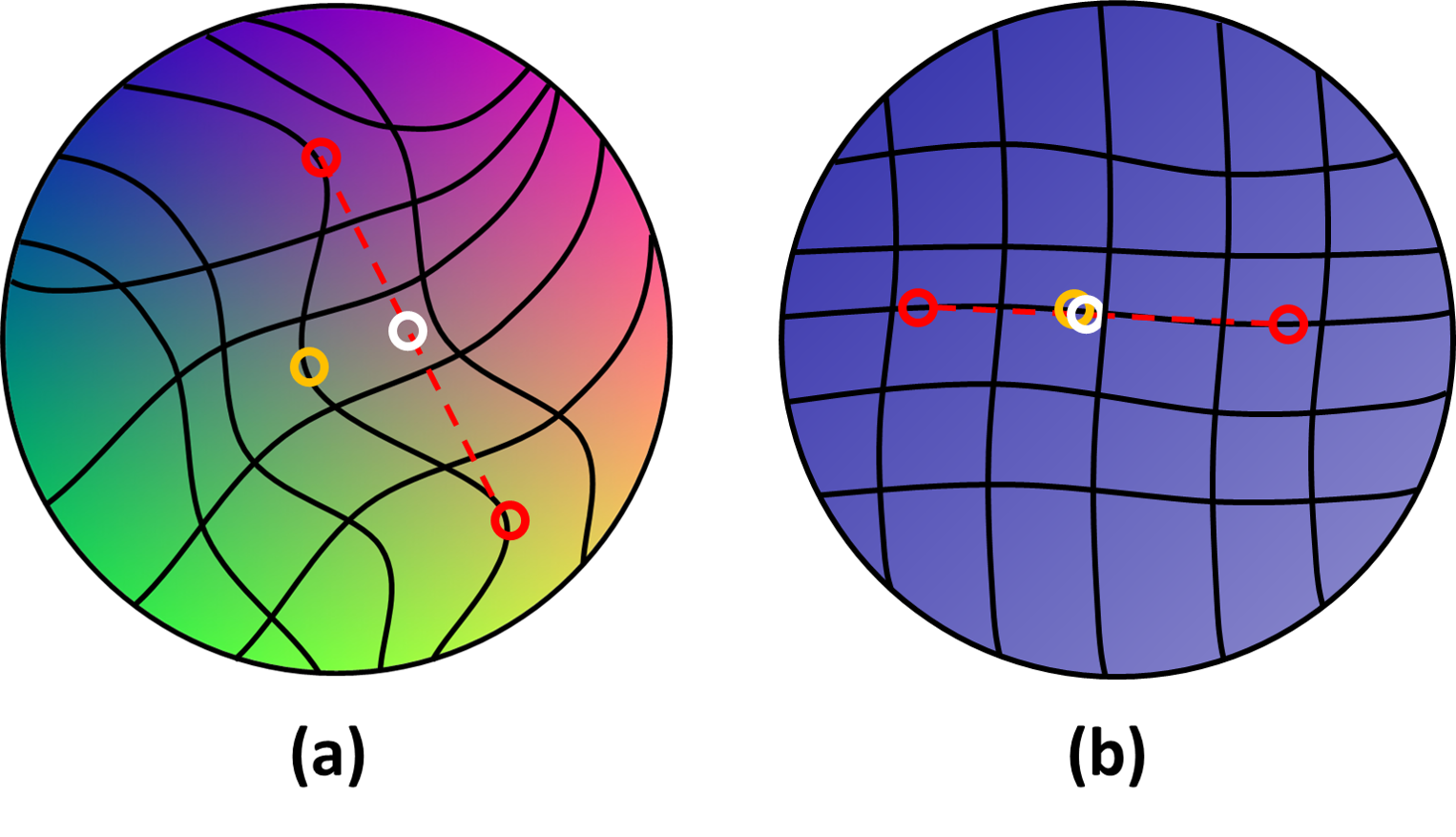}
        \vspace{-0.5cm}
	\caption{The grid geometry with an emotional basis in the embedding space. Embedding space of (a) conventional models and (b) the proposed method. Two red points denote neutral and certain emotion. The yellow and white points are the actual intermediate emotion and the linear interpolated emotion from the two red points, respectively.}
	\label{fig:maniflod}
 	\vspace{-0.5cm}
\end{figure}
\begin{figure*}[t!]
	\centering
	\includegraphics[width=0.8\linewidth]{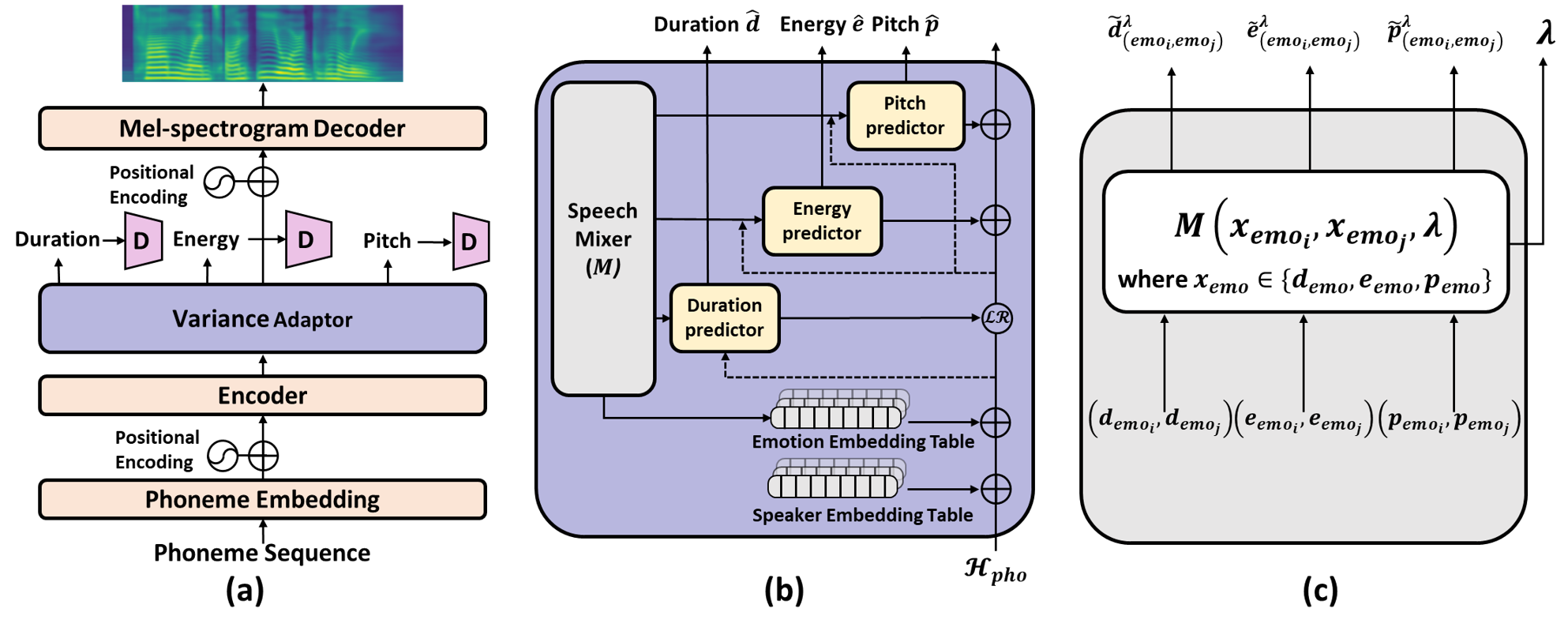}
        \vspace{-0.5cm}
	\caption{ Overall Architecture of proposed model. $\mathcal{H}_{pho}$ is a hidden phoneme embedding and $\lambda$ is an interpolation weight. (a) emotional speech synthesis framework based on Fastspeech2 \cite{3_ren2020fastspeech} (b) variance adaptor (c) speech mixer $M$ which is generating mixed pitch $\tilde{p}_{(emo_i, emo_j)}^\lambda$, duration $\tilde{d}_{(emo_i, emo_j)}^\lambda$, and energy $\tilde{e}_{(emo_i, emo_j)}^\lambda$}
	\vspace{-0.5cm}
    \label{fig:architecture}
\end{figure*}
To improve fine-grained expression, there are attempts to control an emotion intensity \cite{9_li2021controllable,10_li2021controllable,11_um2020emotional,12_im2022emoq,zhu2019controlling,lei2021fine,lei2022msemotts, huprosodybert, zhou2022speech}, not the categorical emotion of speech. \cite{9_li2021controllable, 10_li2021controllable} introduced models to reflect detailed emotional expression by adjusting emotion strength with controllable parameter. \cite{11_um2020emotional} proposed a method to control the intensity of emotions using non-linear interpolation from categorical emotion embedding space. \cite{12_im2022emoq} controlled fine-grained emotion intensity by conducting distance-based intensity quantization. 
\cite{zhu2019controlling,lei2021fine,lei2022msemotts} suggested studies of emotion intensity control with ranking functions and the proposed method is only applicable for a single speaker dataset. 
\cite{huprosodybert} introduced a self-supervised learning for prosody representations. And \cite{zhou2022speech} proposed a method for generating speech with a mixture of emotions.  

Even though previous works have proposed controllable emotional intensity models, there are two limitations. First, it is difficult to synthesize speech by controlling the emotion space as desired. Conventional emotional TTS models find the emotion embedding vector for discretized intervals and utilize the vector to synthesize emotion speech. As shown in Fig. \ref{fig:maniflod}(a), an embedding space is entangled not only with various emotions but also with other features, like speaker identity, pitch or linguistic information. Accordingly, the grid geometry from the perspective of the emotional basis may form a valley-shaped grid as shown in Fig. \ref{fig:maniflod}(a). Due to the valley-shaped gird in the embedding space, linearity for emotions cannot be guaranteed, and it is hard to control emotions as desired. For example, suppose you want to find an intermediate emotion (see the yellow point in Fig. \ref{fig:maniflod}) from two certain emotions (see red points in Fig. \ref{fig:maniflod}). If the embedding space consists of the non-uniform grid as shown in Fig. \ref{fig:maniflod}(a), an emotion predicted by interpolation models is far from the actual intermediate emotion (see the white point in Fig. \ref{fig:maniflod}). Accordingly, the interpolated emotional speech may be synthesized differently than desired. On the other hand, as shown in Fig. \ref{fig:maniflod}(b), the predicted emotion is located close to the actual intermediate emotion if the embedding space is disentangled, so that the desired speech could be synthesized. Second, it is difficult to guarantee the naturalness for intermediate emotional speech, because there are no loss functions or constraints to improve the naturalness. Because of the limitations mentioned above, it is a challenging task to generate the desired speech from the grid of non-uniform emotional latent spaces.

In this work, we propose a method to continuously control the intensity of emotion using semi-supervised learning. In order to learn the speech of intermediate emotions, we propose a novel speech mixer, an augmentation method to interpolate emotion labels and corresponding speech components (pitch, duration and energy). Since the proposed model is directly trained with low-level elements, more fine-grained embedding space can be constructed compared to the conventional emotion latent space. As shown in Fig. \ref{fig:maniflod}(b), the emotion embedding space is not corrupted by other features like speaker and linguistic. 
In addition, a discriminator is applied to the variance adaptor controlling duration, pitch and energy so that the model generates a more realistic low-level element sequences \cite{13_lee2021multi}. 

Contributions in this study are as follows.

\begin{itemize}
    \item By using a novel low-level data mixer to generate intermediate emotion points, the proposed model trained with semi-supervised learning can generate emotional speech with a continuous intensity value.
    \item By applying a discriminator to the variance adaptor, the mel-spectrogram can be generated well without prediction loss.
\end{itemize}

The synthesized speech samples are available at \url{https://tinyurl.com/2p8vdcnd}

\begin{table*}[t!]
\caption{Results of emotion intensity recognition and speech quality evaluation. (i) Emotion Intensity Recognition is the recognition accuracy between two speech samples of different intensity. (ii) Speech Quality Evaluation denotes qualitative metric (MOS) and quantitative metrics (MCD and F0 RMSE). MOS scores are presented with 95$\%$ confidence intervals. MCD and F0 RMSE are evaluated for categorical emotion speech with ground-truth. A to E represent emotional intensity, A=0.00, B=0.25, C=0.50, D=0.75, and E=1.00.}
\centering
\vspace{-0.4cm}
\begin{tabular}{c c cccc ccc}
\noalign{\smallskip}\noalign{\smallskip}\hline\hline
\multirow{2}{4em}{Emotion} & \multirow{2}{4em}{Method} & \multicolumn{4}{c}{(i) Emotion Intensity Recognition [\%]} & \multicolumn{3}{c}{(ii) Speech Quality Evaluation} \\
\cmidrule(lr){3-6} \cmidrule(lr){7-9}
 &  & A $<$ B & B $<$ C & C $<$ D & D $<$ E & MOS $\uparrow$ & MCD $\downarrow$ & F0 RMSE $\downarrow$ \\
\hline
\multirow{2}{*}{(a) Happy}      & Conventional \cite{11_um2020emotional} & 44.727     & 43.636       & 50.909       & 47.636 & 3.528$\pm$0.050 & 5.516 & 103.764 \\
                            &  Proposed  & \bf{57.455} & \bf{58.909} & \bf{58.182} & \bf{58.182}  & \bf{3.594$\pm$0.047} & \bf{5.478} & \bf{86.154} \\
\hline
\multirow{2}{*}{(b) Sad}        & Conventional \cite{11_um2020emotional}  & 46.545     & 44.364       & 40.364       & 44.364 & 3.509$\pm$0.052 & 5.691 & 82.523 \\
                            & Proposed  & \bf{57.818} & \bf{55.273} & \bf{59.636} & \bf{58.182}   & \bf{3.654$\pm$0.045} & \bf{5.470} & \bf{76.468} \\
\hline
\multirow{2}{*}{(c) Angry}      & Conventional \cite{11_um2020emotional} & 48.364     & 47.646       & 45.455       & 45.091 & 3.494$\pm$0.050 & 5.796 & 100.978 \\
                            & Proposed  & \bf{61.818} & \bf{66.545} & \bf{58.182} & \bf{56.727}  &\bf{3.520$\pm$}0.049 & \bf{5.365} & \bf{82.222} \\
\hline
\multirow{2}{*}{(d) Surprise}   & Conventional \cite{11_um2020emotional} & 42.909     & 40.000       & 52.000       & 48.364  & 3.527$\pm$0.051 & 5.280 & 108.823 \\
                            & Proposed  & \bf{62.545} & \bf{64.364} & \bf{63.636} & \bf{58.182}  & \bf{3.659$\pm$0.046} & \bf{5.159} & \bf{84.793} \\
\hline
\hline
\end{tabular}
\vspace{-0.4cm}
\label{tbl:result}
\end{table*}

\vspace{-0.1cm}
\section{Method}
\label{sec:method}

The overall architecture of the proposed model is shown in Fig. \ref{fig:architecture}. Fastspeech2 \cite{3_ren2020fastspeech} is used to generate a mel-spectrogram from the phoneme sequence. We propose a speech mixer $M$ to generate pseudo-labels $\tilde{x}$ reflecting intermediate emotion intensities in a variance adapter. The speech mixer $M$ generates an intermediate low-level elements like pitch $p$, duration $d$, and energy $e$. Also, discriminators $D$ is applied to the predicted elements for improving naturalness.

\subsection{Speech Mixer}
\label{ssec:speech_mixer}
A speech mixer $M$ generates interpolated pseudo-labels $\tilde{x}$ for intermediate emotion intensities. In order to interpolate any two emotions $(emo_i, emo_j)$, emotion speech pair $(S_{emo_i}, S_{emo_j})$ should be sampled from different emotion categories $\mathbb{E} = \{emo_1, emo_2, ..., emo_K\}$ where $K$ denotes the number of emotions. 
In this paper, we used $K=5$ and categorical emotions include neutral, happy, sad, angry, and surprise. 
Its sampling function $F$ can be represented by
\begin{equation*}\label{eq:sampling_function}
S_{emo_j} = F(S_{emo_i}).
\end{equation*}
The emotion speech pair are sampled as follows
\begin{gather*}\label{eq:sampling_rule}
(emo_i = neutral, ~emo_j \in \mathbb{E} \setminus \{neutral\}), \\
(\text{resp.}~~~ (emo_i \in \mathbb{E} \setminus \{neutral\}, ~emo_j = neutral) ~~~).
\end{gather*}

To generate a pseudo-label $\tilde{x}$, sampled pair $(S_{emo_i}, S_{emo_j})$ is converted into phoneme-level averaged values, so that the same sentences have the same length of pitch  $(p_{emo_i}, p_{emo_j})$, duration $(d_{emo_i}, d_{emo_j})$ and energy $(e_{emo_i}, e_{emo_j})$. Then speech mixer $M$ generates pseudo-labels $\tilde{x}_{(emo_i, emo_j)}^\lambda$ for intermediate intensity of emotional speech, given by
\begin{align*}\label{mixer}
M(x_{emo_i}, x_{emo_j}, \lambda)    &= g(\lambda x_{emo_i} + (1-\lambda) x_{emo_j} )\\
                                    &= \tilde{x}_{(emo_i, emo_j)}^\lambda,
\end{align*}
where $x_{emo} \in \{p_{emo}, d_{emo}, e_{emo}\}$ and $\lambda$ denotes an interpolation weight. $g(\cdot)$ denotes floor function if $x_{emo} = d_{emo}$ else identity function. Specifically, the interpolation weight $\lambda$ is randomly selected from beta distribution $\beta(0.5, 0.5)$.
For notation simplicity, we denote $x_{emo} = x$ and $\tilde{x}_{(emo_i, emo_j)}^\lambda = \tilde{x}$.

\subsection{Generator}
\label{ssec:generator}

As shown in Fig. \ref{fig:architecture}(a), we use FastSpeech2 \cite{3_ren2020fastspeech}, which consists of a variance adapter, phoneme-encoder, and decoder. The phoneme encoder receives a phoneme sequence as an input and outputs an embedding vector. After adding a positional encoding to the embedding vector, the encoder produces a hidden phoneme embedding $\mathcal{H}_{pho}$.

Speaker and emotion Look-Up Tables (LUTs) are introduced to extend the existing variance adapter to a multi-speaker setting like Fig. \ref{fig:architecture}(b). The speaker LUT is assigned to each speaker and trained to suit the speaker. The emotion LUTs also are optimized according to the emotion labels.
These speaker and emotion labels are obtained from the dataset, and the details of dataset are in Section 3.1. 
To optimize the phoneme embedding $\mathcal{H}_{pho}$, the speaker and the emotion LUTs, loss functions for training each low-level element are described as follows.

Loss of duration $\mathcal{L}_d$ consists of mean-square error (MSE) of logarithm function such that 
\begin{equation}
\mathcal{L}_{d} = \mathop{\mathbb{E}[|| \log(d+1) - \log(\hat{d})||_{2}]},
\label{eq:d_loss}
\end{equation}
where $d$ and $\hat{d}$ are a phoneme-level duration and its predicted value from a duration predictor, respectively. 
Similar to loss of duration $\mathcal{L}_d$, loss functions of pitch $\mathcal{L}_p$ and energy $\mathcal{L}_e$ are formulated as MSE, given by
\begin{equation}
\mathcal{L}_{p} = \mathop{\mathbb{E}[|| p - \hat{p}||_{2}]}, \quad \mathcal{L}_{e} = \mathop{\mathbb{E}[|| e - \hat{e}||_{2}]},
\label{eq:pe_loss}
\end{equation}
where $p$ and $e$ are labels of pitch and energy, respectively. $\hat{p}$ and $\hat{e}$ denote predicted values from pitch and energy predictors. 
For Eqs. \eqref{eq:d_loss} and \eqref{eq:pe_loss}, labels $x \in \{d, p, e\}$ can be replaced with pseudo-labels $\tilde{x} \in \{\tilde{d}, \tilde{p}, \tilde{e}\}$.

\subsection{Discriminator}
\label{ssec:subhead}
Low-level elements generated by the speech mixer do not exist a corresponding speech ground-truth, so it is difficult to guarantee naturalness. Adversarial training scheme is conducted to help the variance adaptor generate more realistic pitch, duration and energy sequences. We adopt the least squares GAN \cite{15_mao2017least} loss for training our proposed model. Discriminators are shown as D in Fig. \ref{fig:architecture}(a), which are trained adversarially on the predicted pitch $\hat{p}$, duration $\hat{d}$, and energy $\hat{e}$ from the variance adapter. The adversarial loss $\mathcal{L}^{adv}_{x} $ is as follows:
\begin{equation}\label{discriminator_adv_loss}
\mathcal{L}^{adv}_{x} = \mathop{\mathbb{E}[(x - 1)^2] + \mathbb{E}[( \tilde{x} )^2]}
\end{equation}

\subsection{Training Objectives}
\label{ssec:training_obj}
Network training consists of two phases; (1) learning categorical emotion using the original dataset $x$, (2) learning intermediate emotion using pseudo-label data $\tilde{x}$ generated from a speech mixer $M$. First, when the model is trained with a categorical dataset $x$, Eqs. \eqref{eq:d_loss} and \eqref{eq:pe_loss} are used, and mean-absolute error (MAE) loss is also computed between a ground-truth mel-spectrogram $y$ and predicted mel-spectrogram $\hat{y}$, given by 
\begin{equation}\label{mel_loss}
\mathcal{L}_{mel} = \mathop{\mathbb{E}[|| y - \hat{y}||_{1}]}.
\end{equation}
So, categorical loss is defined as
\begin{equation*}\label{total_cate_loss}
\mathcal{L}_{categorical} = \mathcal{L}_{mel} + \mathcal{L}_{p} + \mathcal{L}_d +\mathcal{L}_e.
\end{equation*}
Second, when the network is trained with intermediate emotion $\tilde{x}$ generated from a speech mixer $M$, MSE losses are used similarly to a categorical loss $\mathcal{L}_{categorical}$. However, the adversarial loss is additionally applied to each pseudo-label $\tilde{x}$, instead of Eq. \eqref{mel_loss}, given by
\begin{equation}\label{adv_loss}
 \mathcal{L}_{adv} = \mathcal{L}^{adv}_{p} + \mathcal{L}^{adv}_{d} + \mathcal{L}^{adv}_{e}.
\end{equation}
So, intermediate loss is defined as
\begin{equation*}\label{intermediate_loss}
 \mathcal{L}_{intermediate}=\mathcal{L}_{adv}+\mathcal{L}_{\tilde{p}}+\mathcal{L}_{\tilde{d}}+\mathcal{L}_{\tilde{e}}
\end{equation*}
Finally, total training loss consists of categorical loss and intermediate loss.
as follows 
\begin{equation*}\label{total_mix_loss}
\mathcal{L}_{total}=\mathcal{L}_{categorical} + \mathcal{L}_{intermediate}
\end{equation*}

\vspace{-0.2cm}
\section{Experiments and Results}
\label{sec:experiments}

\begin{table*}[t!]
\caption{ Ablation study of discriminator and interpolation weight $\lambda$. Scores are average of all emotions. Discrete means that the data mixing ratio is randomly selected from 0, 0.5, or 1.0. Uniform means that the ratio is sampled from the uniform (0, 1) distribution.}
\vspace{-0.4cm}
\centering
\begin{tabular}{ccccccccccc}
\noalign{\smallskip}\noalign{\smallskip}\hline\hline
\multirow{2}{*}{Proposed} & \multirow{2}{*}{Weight $\lambda$} & \multicolumn{4}{c}{(i) Emotion Intensity Recognition [\%]} & \multicolumn{3}{c}{(ii) Speech Quality Evaluation} \\
\cmidrule(lr){3-6} \cmidrule(lr){7-9}
 &  &   A $<$ B & B $<$ C & C $<$ D & D $<$ E   & MOS $\uparrow$ & MCD $\downarrow$ & F0 RMSE $\downarrow$  \\
\hline
(a) w/o discriminator &  Beta & 46.182 & 44.455 & 43.636 & 44.455  & 3.597$\pm$0.023 & 5.415  & \bf{77.133}  \\ \hline
\multirow{3}{*}{(b) w/ discriminator}  &  Discrete  & 54.091 & 50.818 & 50.818 & 53.273   & 3.602$\pm$0.023  & \bf{5.337}  &   79.359  \\
 &  Uniform & 43.455   & 41.455 & 40.000 & 41.455   & 3.589$\pm$0.024 & 5.367  & 79.562  \\
 &  Beta & \bf{59.909} & \bf{61.273} & \bf{59.909} & \bf{57.818} & \bf{3.607$\pm$0.045} & 5.362 & 82.409 \\  

\hline
\hline
\end{tabular}

\label{tbl:ablation}
\vspace{-0.5cm}
\end{table*}

\subsection{Dataset}
\label{ssec:dataset}
We used Emotional Speech Database (ESD) \cite{16_zhou2021seen} for multi-speaker models. The ESD covers five emotions (neutral, happy, angry, sad and surprise) and comprises of 350 parallel utterances from 10 native English speakers and 10 native Chinese speakers. We only used the English dataset with all emotions for training and evaluation. It is split into train, validation and test and 1000 sentences are used as validation and test set to evaluate the performance. 

\vspace{-0.1cm}
\subsection{Training Details}
\label{ssec:training_details}
We transformed the raw waveform into mel-spectrogram and set hop size to 256 and mel bins to 80. Montreal forced alignment \cite{mcauliffe2017montreal} of version 1.1.4 was used to extract the phoneme duration. We used pretrained Hifi-gan \cite{kong2020hifi} universal version as a vocoder and trained the rest parts from scratch. 
We trained Adam with $\beta_{1}$ = 0.9,  $\beta_{2}$ = 0.98, $\epsilon = 10^{-9}$ and set learning rate to $ 10^{-5}$. 
The model was trained using 64 batch size with 800k steps for training until convergence and the number of trainable parameters is about 3.5M. All experiments were carried out on a single RTX2080 GPU and took about 7days for training.

\vspace{-0.1cm}
\subsection{Model Performance}
\label{ssec:model_performance}
We conducted a preference test using Amazon Mechanical Turk to assess emotion intensity recognition. 11 sentences were randomly sampled per emotion, and 220 participants were involved.
First, the raters listen to the same speaker and speech uttered with a neutral emotion, and speech uttered with a specific emotion as a reference. Then, two sentences uttered with different intensities are given, and among the two sentences, raters should select the one with the stronger emotion. A specific emotion is one of four emotions like happy, sad, angry, or surprise, and 4 intensity types were tested. There are 4 types such as (0.0 vs 0.25), (0.25 vs 0.5), (0.5 vs 0.75), and (0.75 vs 1.0). For speech quality evaluation, mean opinion score (MOS) \cite{19_streijl2016mean} was measured through a questionnaire to verify the speech naturalness. For categorical emotional speech, mel cepstral distortion (MCD) \cite{18_kubichek1993mel} and F0 root mean square error (F0 RMSE) were computed for quantitative evaluation. Conventional method \cite{11_um2020emotional} controls emotion intensity through non-linear interpolation based on GST \cite{4_wang2018style}. As shown in Table \ref{tbl:result}, the proposed method outperforms the conventional model \cite{11_um2020emotional} in all metrics. Specifically, Table \ref{tbl:result}(i) shows that our proposed method achieves the best accuracy for all intensity types. This indicates that the proposed model can synthesize speech well according to the given intensity scale. In addition, for speech quality evaluation, the proposed method showed better performance than the conventional model \cite{11_um2020emotional} in all emotions as shown in Table \ref{tbl:result}(ii).

\begin{figure}[t!]
	\centering
	\includegraphics[width=0.7\linewidth]{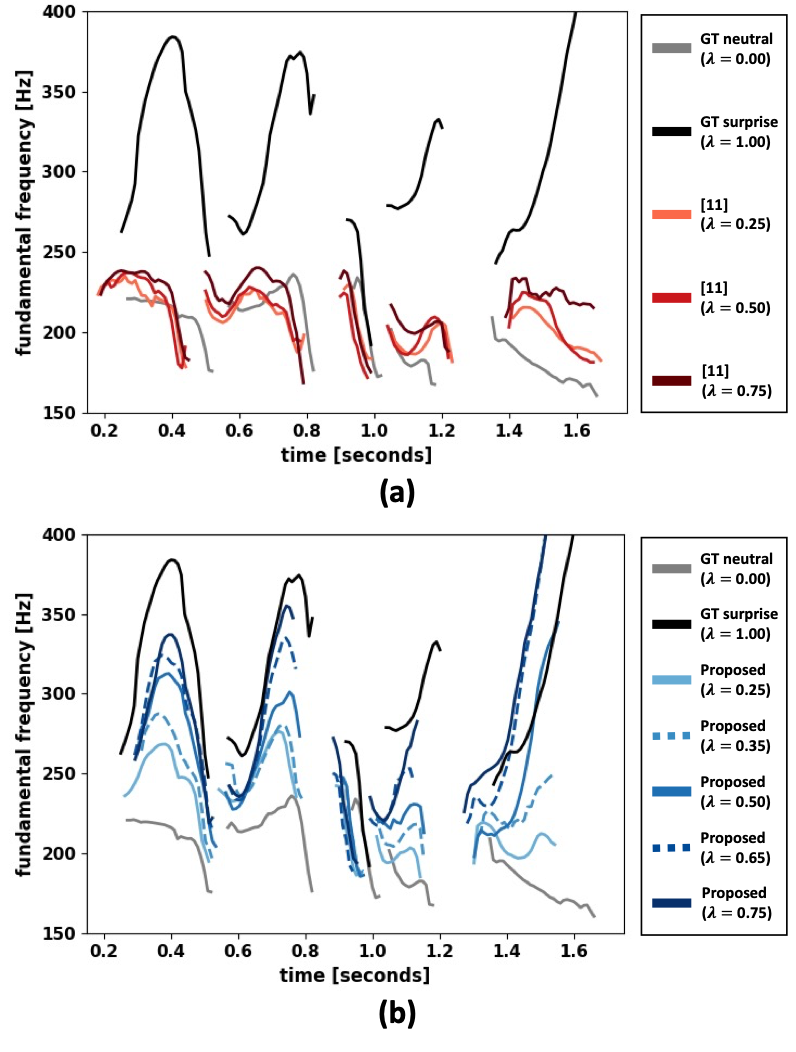}
 	\vspace{-0.5cm}
	\caption{Plotting pitch contours of (a) conventional method \cite{11_um2020emotional} and (b) proposed method according to emotional intensity.}
	\label{fig:pitch}
 	\vspace{-0.4cm}
\end{figure}

\vspace{-0.1cm}
\subsection{Ablation Study}
\label{ssec:ablation_study}
We conducted an ablation study to validate the effectiveness of the discriminator. In the proposed model w/o discriminator at Table \ref{tbl:ablation}(a), all types of emotion intensity accuracy decreased compared to the model w/ discriminator when $\lambda$ distribution is beta (see Table \ref{tbl:ablation}(i)). However, for the F0 RMSE metric as shown in Table \ref{tbl:ablation}(ii), the model w/o discriminator represented better performance than w/ discriminator since the model w/o discriminator was only optimized to minimize regression losses related to labels and pseudo-labels. In addition, another ablation study was conducted for different interpolation weight distributions of speech mixer $M$. We compared discrete and uniform distributions as interpolation weight $\lambda$. Discrete distribution means that the mixing ratio $\lambda$ is randomly sampled from among 0, 0.5, and 1.0. And uniform means that the ratio $\lambda$ is sampled from the uniform distribution $\mathcal{U}(0, 1)$. The proposed model trained with the speech mixer using beta distribution $\beta(0.5, 0.5)$ shows the best performance of the emotion intensity recognition as shown in Table \ref{tbl:ablation}(a)(i).
However, the model with discrete distribution achieved the best MCD and F0 RMSE scores except w/o discriminator (see Table \ref{tbl:ablation}(b)(ii)). The model trained with the discrete distribution can frequently encounter categorical labels and be optimized, thus the quantitative metrics are minimized.

\subsection{Plotting pitch contours of samples}
\label{ssec:plotting}

Synthesized speech samples of the proposed model and conventional model \cite{11_um2020emotional} were analyzed. The pitch contour was plotted for the same speaker and sentence as shown in Fig. \ref{fig:pitch}. The pitch contour of the proposed model dynamically changed according to the emotional intensity $\lambda$. However, the conventional model \cite{11_um2020emotional} showed similar pitch contours despite the intensity $\lambda$ being modified from 0.25 to 0.75. In particular, the proposed model can synthesize the speech at any emotional intensity (see the dashed line in Fig. \ref{fig:pitch}(b)) although the conventional model \cite{11_um2020emotional} cannot (see Fig. \ref{fig:pitch}(a)). It means that the pitch sequences can be controlled by selecting the desired intensity with any continuous value. Thus, we confirmed that our proposed model can dynamically adjust the intensity of emotions.

\vspace{-0.2cm}
\section{Conclusion}
\label{sec:conclusion}
Improving expression in speech synthesis is very important but challenging task. In particular, for supervised learning, labeling a dataset that can control the emotions of speech is a laborious and difficult task. Therefore, we proposed a model that can control the emotional intensity with continuous value using semi-supervised learning. Intermediate low-level elements are generated for a categorical emotional speech dataset, and it is used as a pseudo-label for network learning. This study has a limitation in that the parallel expressive data corpus is necessary. The ground-truth mel-spectrogram does not exist in the pseudo-labels, so a discriminator is used to supplement it. The proposed model through experiments showed superior performance in emotional intensity control and naturalness.

\vspace{-0.2cm}
\section{Acknowledgement}
\label{sec:acknowledgement}
This work was partly supported by Institute of Information \& communications Technology Planning \& Evaluation (IITP) grant funded by the Korea government(MSIT) (No.2022-0-00320, Artificial intelligence research about cross-modal dialogue modeling for one-on-one multi-modal interactions, 50\%) and (No. 2022-0-00641, XVoice: Multi-Modal Voice Meta Learning, 50\%)

\bibliographystyle{IEEEtran}
\bibliography{mybib}

\begin{thebibliography}{10}
\providecommand{\url}[1]{#1}
\csname url@samestyle\endcsname
\providecommand{\newblock}{\relax}
\providecommand{\bibinfo}[2]{#2}
\providecommand{\BIBentrySTDinterwordspacing}{\spaceskip=0pt\relax}
\providecommand{\BIBentryALTinterwordstretchfactor}{4}
\providecommand{\BIBentryALTinterwordspacing}{\spaceskip=\fontdimen2\font plus
\BIBentryALTinterwordstretchfactor\fontdimen3\font minus
  \fontdimen4\font\relax}
\providecommand{\BIBforeignlanguage}[2]{{%
\expandafter\ifx\csname l@#1\endcsname\relax
\typeout{** WARNING: IEEEtran.bst: No hyphenation pattern has been}%
\typeout{** loaded for the language `#1'. Using the pattern for}%
\typeout{** the default language instead.}%
\else
\language=\csname l@#1\endcsname
\fi
#2}}
\providecommand{\BIBdecl}{\relax}
\BIBdecl

\bibitem{1_tan2022naturalspeech}
X.~Tan, J.~Chen, H.~Liu, J.~Cong, C.~Zhang, Y.~Liu, X.~Wang, Y.~Leng, Y.~Yi,
  L.~He \emph{et~al.}, ``Naturalspeech: End-to-end text to speech synthesis
  with human-level quality,'' \emph{arXiv preprint arXiv:2205.04421}, 2022.

\bibitem{2_kim2021conditional}
J.~Kim, J.~Kong, and J.~Son, ``Conditional variational autoencoder with
  adversarial learning for end-to-end text-to-speech,'' in \emph{International
  Conference on Machine Learning}.\hskip 1em plus 0.5em minus 0.4em\relax PMLR,
  2021, pp. 5530--5540.

\bibitem{3_ren2020fastspeech}
Y.~Ren, C.~Hu, X.~Tan, T.~Qin, S.~Zhao, Z.~Zhao, and T.~Liu, ``Fastspeech 2:
  Fast and high-quality end-to-end text to speech,'' in \emph{International
  Conference on Learning Representations, {ICLR}}, 2021.

\bibitem{4_wang2018style}
Y.~Wang, D.~Stanton, Y.~Zhang, R.-S. Ryan, E.~Battenberg, J.~Shor, Y.~Xiao,
  Y.~Jia, F.~Ren, and R.~A. Saurous, ``Style tokens: Unsupervised style
  modeling, control and transfer in end-to-end speech synthesis,'' in
  \emph{International Conference on Machine Learning}.\hskip 1em plus 0.5em
  minus 0.4em\relax PMLR, 2018, pp. 5180--5189.

\bibitem{5_wu2019end}
P.~Wu, Z.~Ling, L.~Liu, Y.~Jiang, H.~Wu, and L.~Dai, ``End-to-end emotional
  speech synthesis using style tokens and semi-supervised training,'' in
  \emph{2019 Asia-Pacific Signal and Information Processing Association Annual
  Summit and Conference (APSIPA ASC)}.\hskip 1em plus 0.5em minus 0.4em\relax
  IEEE, 2019, pp. 623--627.

\bibitem{6_cai2021emotion}
X.~Cai, D.~Dai, Z.~Wu, X.~Li, J.~Li, and H.~Meng, ``Emotion controllable speech
  synthesis using emotion-unlabeled dataset with the assistance of cross-domain
  speech emotion recognition,'' in \emph{ICASSP 2021-2021 IEEE International
  Conference on Acoustics, Speech and Signal Processing (ICASSP)}.\hskip 1em
  plus 0.5em minus 0.4em\relax IEEE, 2021, pp. 5734--5738.

\bibitem{7_lee2017emotional}
Y.~Lee, A.~Rabiee, and S.-Y. Lee, ``Emotional end-to-end neural speech
  synthesizer,'' \emph{arXiv preprint arXiv:1711.05447}, 2017.

\bibitem{8_tits2019exploring}
N.~Tits, K.~El~Haddad, and T.~Dutoit, ``Exploring transfer learning for low
  resource emotional tts,'' in \emph{Proceedings of SAI Intelligent Systems
  Conference}.\hskip 1em plus 0.5em minus 0.4em\relax Springer, 2019, pp.
  52--60.

\bibitem{9_li2021controllable}
T.~Li, S.~Yang, L.~Xue, and L.~Xie, ``Controllable emotion transfer for
  end-to-end speech synthesis,'' in \emph{2021 12th International Symposium on
  Chinese Spoken Language Processing (ISCSLP)}.\hskip 1em plus 0.5em minus
  0.4em\relax IEEE, 2021, pp. 1--5.

\bibitem{10_li2021controllable}
T.~Li, X.~Wang, Q.~Xie, Z.~Wang, and L.~Xie, ``Controllable cross-speaker
  emotion transfer for end-to-end speech synthesis,'' \emph{arXiv preprint
  arXiv:2109.06733}, 2021.

\bibitem{11_um2020emotional}
S.-Y. Um, S.~Oh, K.~Byun, I.~Jang, C.~Ahn, and H.-G. Kang, ``Emotional speech
  synthesis with rich and granularized control,'' in \emph{ICASSP 2020-2020
  IEEE International Conference on Acoustics, Speech and Signal Processing
  (ICASSP)}.\hskip 1em plus 0.5em minus 0.4em\relax IEEE, 2020, pp. 7254--7258.

\bibitem{12_im2022emoq}
C.-B. Im, S.-H. Lee, S.-B. Kim, and S.-W. Lee, ``Emoq-tts: Emotion intensity
  quantization for fine-grained controllable emotional text-to-speech,'' in
  \emph{ICASSP 2022-2022 IEEE International Conference on Acoustics, Speech and
  Signal Processing (ICASSP)}.\hskip 1em plus 0.5em minus 0.4em\relax IEEE,
  2022, pp. 6317--6321.

\bibitem{zhu2019controlling}
X.~Zhu, S.~Yang, G.~Yang, and L.~Xie, ``Controlling emotion strength with
  relative attribute for end-to-end speech synthesis,'' in \emph{2019 IEEE
  Automatic Speech Recognition and Understanding Workshop (ASRU)}.\hskip 1em
  plus 0.5em minus 0.4em\relax IEEE, 2019, pp. 192--199.

\bibitem{lei2021fine}
Y.~Lei, S.~Yang, and L.~Xie, ``Fine-grained emotion strength transfer, control
  and prediction for emotional speech synthesis,'' in \emph{2021 IEEE Spoken
  Language Technology Workshop (SLT)}.\hskip 1em plus 0.5em minus 0.4em\relax
  IEEE, 2021, pp. 423--430.

\bibitem{lei2022msemotts}
Y.~Lei, S.~Yang, X.~Wang, and L.~Xie, ``Msemotts: Multi-scale emotion transfer,
  prediction, and control for emotional speech synthesis,'' \emph{IEEE/ACM
  Transactions on Audio, Speech, and Language Processing}, vol.~30, pp.
  853--864, 2022.

\bibitem{huprosodybert}
Y.~Hu, C.~Zhang, J.~Shi, J.~Lian, M.~Ostendorf, and D.~Yu, ``Prosodybert:
  Self-supervised prosody representation for style-controllable tts.''

\bibitem{zhou2022speech}
K.~Zhou, B.~Sisman, R.~Rana, B.~W. Schuller, and H.~Li, ``Speech synthesis with
  mixed emotions,'' \emph{IEEE Transactions on Affective Computing}, 2022.

\bibitem{13_lee2021multi}
S.-H. Lee, H.-W. Yoon, H.-R. Noh, J.-H. Kim, and S.-W. Lee, ``Multi-spectrogan:
  High-diversity and high-fidelity spectrogram generation with adversarial
  style combination for speech synthesis,'' in \emph{Proceedings of the AAAI
  Conference on Artificial Intelligence}, vol.~35, no.~14, 2021, pp.
  13\,198--13\,206.

\bibitem{15_mao2017least}
X.~Mao, Q.~Li, H.~Xie, R.~Y. Lau, Z.~Wang, and S.~Paul~Smolley, ``Least squares
  generative adversarial networks,'' in \emph{Proceedings of the IEEE
  international conference on computer vision}, 2017, pp. 2794--2802.

\bibitem{16_zhou2021seen}
K.~Zhou, B.~Sisman, R.~Liu, and H.~Li, ``Seen and unseen emotional style
  transfer for voice conversion with a new emotional speech dataset,'' in
  \emph{ICASSP 2021-2021 IEEE International Conference on Acoustics, Speech and
  Signal Processing (ICASSP)}.\hskip 1em plus 0.5em minus 0.4em\relax IEEE,
  2021, pp. 920--924.

\bibitem{mcauliffe2017montreal}
M.~McAuliffe, M.~Socolof, S.~Mihuc, M.~Wagner, and M.~Sonderegger, ``Montreal
  forced aligner: Trainable text-speech alignment using kaldi.'' in
  \emph{Interspeech}, vol. 2017, 2017, pp. 498--502.

\bibitem{kong2020hifi}
J.~Kong, J.~Kim, and J.~Bae, ``Hifi-gan: Generative adversarial networks for
  efficient and high fidelity speech synthesis,'' \emph{Advances in Neural
  Information Processing Systems}, vol.~33, pp. 17\,022--17\,033, 2020.

\bibitem{19_streijl2016mean}
R.~C. Streijl, S.~Winkler, and D.~S. Hands, ``Mean opinion score (mos)
  revisited: methods and applications, limitations and alternatives,''
  \emph{Multimedia Systems}, vol.~22, no.~2, pp. 213--227, 2016.

\bibitem{18_kubichek1993mel}
R.~Kubichek, ``Mel-cepstral distance measure for objective speech quality
  assessment,'' in \emph{Proceedings of IEEE pacific rim conference on
  communications computers and signal processing}, vol.~1.\hskip 1em plus 0.5em
  minus 0.4em\relax IEEE, 1993, pp. 125--128.

\end{thebibliography}

\end{document}